\documentclass[10pt,preprint]{aastex}
\usepackage{epsfig,color}
\usepackage{mathrsfs}

\slugcomment{Accepted by ApJ on Jan. 26, 2012}
\shorttitle{Clumpy accretion onto black holes}

\def\mathfrakm{{\mathfrak{M}}}
\def\ergs{${\rm erg~s^{-1}}$}
\def\FR{F_{_R}}
\def\Fphi{F_{\phi}}
\def\mbh{M_{\bullet}}
\def\mathf{{\mathscr F}}
\def\mathn{{\mathscr N}}

\def\kb{k_{\rm B}}

\def\pf{\partial {\mathscr F}}
\def\pp{\partial}
\def\pvphi{\langle v_{\phi}\rangle}
\def\pvr{\langle v_{_R}\rangle}
\def\pvrvz{\langle v_{_R}v_z\rangle}
\def\pvrphi{\langle v_{_R}v_{\phi}\rangle}
\def\pvphiz{\langle v_{\phi}v_z\rangle}
\def\pvvphi{\langle v_{\phi}^2\rangle}
\def\pvvr{\langle v_R^2\rangle}
\def\pvz{\langle v_z\rangle}
\def\pvvz{\langle v_z^2\rangle}

\def\sigmaphi{\sigma_{\phi}}
\def\sigmar{\sigma_{_R}}
\def\sunm{M_\odot}
\def\vr{v_{_R}}
\def\Vr{V_{_R}}
\def\vphi{v_{\phi}}
\def\Vphi{V_{\phi}}

\begin{document}

\title{Clumpy accretion onto black holes. I. Clumpy-ADAF structure and radiation}

\author{Jian-Min Wang\altaffilmark{1,2},
Cheng Cheng\altaffilmark{1}
and Yan-Rong Li\altaffilmark{1}}

\altaffiltext{1}
{Key Laboratory for Particle Astrophysics, Institute of High Energy Physics,
Chinese Academy of Sciences, 19B Yuquan Road, Beijing 100049, China}

\altaffiltext{2}
{National Astronomical Observatories of China, Chinese Academy of Sciences, 20A Datun Road,
Beijing 100020, China}

\begin{abstract}
In this paper, we investigate the dynamics of clumps embedded in and confined by the advection-dominated
accretion
flows (ADAF), in which collisions among the clumps are neglected. We start from the collisionless Boltzmann
equation and assume that interaction between the clumps and the ADAF is responsible for transporting angular
momentum of clumps outward. The inner edge of the clumpy-ADAF is set to be the tidal radius of the clumps. We
consider strong and weak coupling cases, in which the averaged properties of clumps follow the ADAF dynamics
and mainly determined by the black hole potential, respectively. We get the analytical solution
of the dynamics of clumps for the two cases. The velocity dispersion of clumps is one magnitude higher
than the ADAF for the strong coupling case. For the weak coupling case, we find that the mean radial
velocity of clumps is linearly proportional to the coefficient of the drag force. We show
that the tidally disrupted clumps would lead to accumulation of the debris to form a debris disk
in the Shakura-Sunyaev regime. The entire hot ADAF will be efficiently cooled down by photons from
the debris disk, giving rise to collapse of the ADAF and quench the clumpy accretion. Subsequently,
evaporation of the collapsed ADAF drives resuscitate of a new clumpy-ADAF, resulting in an oscillation
of the global clumpy-ADAF. Applications of the present model are briefly discussed to X-ray binaries,
ionization nuclear emission regions (LINERs) and BL Lac objects.
\end{abstract}

\section{Introduction}
Accretion onto black holes is energy sources of various kinds of celestial high energy objects.
Radiation hydrodynamics of the accretion has been established well and known as the standard accretion
disk model (Shakura \& Sunyaev 1973), the slim accretion disk (Abramowicz et al. 1988; Wang \& Zhou 1999;
Wang \& Netzer 2003) and the advection-dominated accretion flows (ADAF) (Narayan \& Yi 1994) in light of
dimensionless accretion rates. These models are widely applied, however, it is not clear yet to what
extend the known models represent a realistic description of the observed phenomena. Moreover, it
should be noted that these models are based on the continuous fluid with radiation fields whereas
the continuous disk is undergoing the thermal, viscosity or photon bubble instabilities. Clearly,
the popular treatment of accretion disks as continuous fluid only holds as a zeroth-order approximation.

It arises from both theoretical and observational motivations that accretion onto black holes
is clumpy rather than homogeneously continuous. Instabilities of the radiation-pressure dominated
regions driven by thermal (Krolik 1998), magneto-rotational (Blaes \& Socrates 2001; 2003) and
photon bubble instabilities (Gammie 1998) create cold clumps in the disk, forming multi-phase
medium around the black hole. As a general case of the simplest version, the two-phase disk-corona
model has been suggested for many years (e.g. Galeev et al. 1979; Haardt \& Maraschi 1993; Mayer
\& Pringle 2007). More generally, clumpy disk has been suggested for many years in light of the
X-ray properties of X-ray binaries and active galactic nuclei (AGNs) (Guilbert \& Rees 1988; Celotti
et al. 1992; Collin-Souffrin et al. 1996; Kuncic et al. 1997; Celotti \& Rees 1999; Yuan 2003;
Lawrence 2011). Recently, low-luminosity
AGNs (LLAGNs) presumed to be powered by the ADAF show components of big blue bumps like
brighter AGNs and quasars (Maoz et al. 2007; but see Ho 2008 for a review), implying that
there are cold matters in the hot flows. There are motivated arguments for the existence of clumpy
disk both in AGNs (Kunzic et al. 1996; Kumar 1999) including low luminosity AGNs (Celotti \& Rees
1999) and X-ray binaries (Malzac \& Celotti 2002; Merloni et al. 2006). Similar to the LLAGNs, some
X-ray binaries show broad K$\alpha$ components in the low states (Miller et al. 2006a,b; Tomsick et
al. 2008; Reis et al. 2009, 2010). Most of previous efforts focus on the internal state of clumps
and their reprocessing properties (Guilbert \& Rees 1988; Celotti et al. 1992; Kuncic et al. 1996, 1997;
Malzac \& Celotti 2002; Merloni et al. 2006), however, dynamics of clumps in disks is insufficiently
understood.

On the other hand, fates of the clumps embedded in accretion flows are poorly known when they are
approaching
the black hole. They would be tidally disrupted by the hole, of which the captured debris is eventually
accreted onto the hole. Unlike the case of black hole capturing stars, the capture rates of clumps
are so fast that the debris of disrupted clumps is accumulating with time. In this paper, we show
that the emission from accretion of debris can efficiently cool the hot ADAF, and leads it to collapse,
quenching the clumpy accretion. Being triggered through $\alpha-$viscosity or evaporation, the collapsed
ADAF (cADAF) revives as a new clumpy-ADAF. This is a cycle between clumpy-ADAF and the cADAF,
which is driven by the clumps. Radiation from the clumpy-ADAF shows interesting temporal properties.

In this paper, we presume that clumpy structure in the ADAF has been formed through some mechanisms
listed above in the ADAF, or produced in the transition regions between the ADAF and the Shakura-Sunyaev
disk. Collisions among the clumps can be neglected in the present case. The goal of the present paper is
to derive the dynamical equations of clumpy-ADAF and we fortunately obtain the analytical solution
of the clump dynamics in the ADAF. We find that the captured clumps will accumulate from the tidally
disruption radius, and the radiation from the debris disk efficiently cool the ADAF.
The presence of the debris disk is driving the global clumpy-ADAF to oscillate.
The model is briefly applied to X-ray binaries and low luminosity AGNs.

\section{Assumptions and dynamical equations}
\subsection{Basic assumptions}
Figure 1 shows the regimes of accretion disk models. We simply refer that slim disks have $\dot{m}\gtrsim 1$,
standard model of Shakura-Sunyaev disk (SSD) works between $ 1\gtrsim \dot{m}\gtrsim \dot{m}_2$, and
accretion flows become advection-dominated when $\dot{m}_2=\alpha_{\rm A}^2\approx 0.1\alpha_{0.3}^2$,
where $\alpha_{0.3}=\alpha_{\rm A}/0.3$ is the viscosity parameter of the ADAF,
$\dot{m}=\dot{M}/\dot{M}_{\rm Edd}$, $\dot{M}$ is the accretion rates,
$\dot{M}_{\rm Edd}=L_{\rm Edd}/\eta c^2=1.39\times 10^{18}~\eta_{0.1}^{-1}m_{\bullet}~{\rm g~s^{-1}}$,
$c$ is the light speed, $\eta_{0.1}=\eta/0.1$ is the radiative efficiency,
$L_{\rm Edd}=1.25\times 10^{38}~m_{\bullet}$\ergs is the Eddington luminosity
and $m_{\bullet}=\mbh/\sunm$ is the black hole mass. When the accretion rates are low enough
($\dot{m}<\dot{m}_1$), flows become a pure ADAF without clumps.

{
\centering
\figurenum{1}
\includegraphics[angle=-90,scale=0.45]{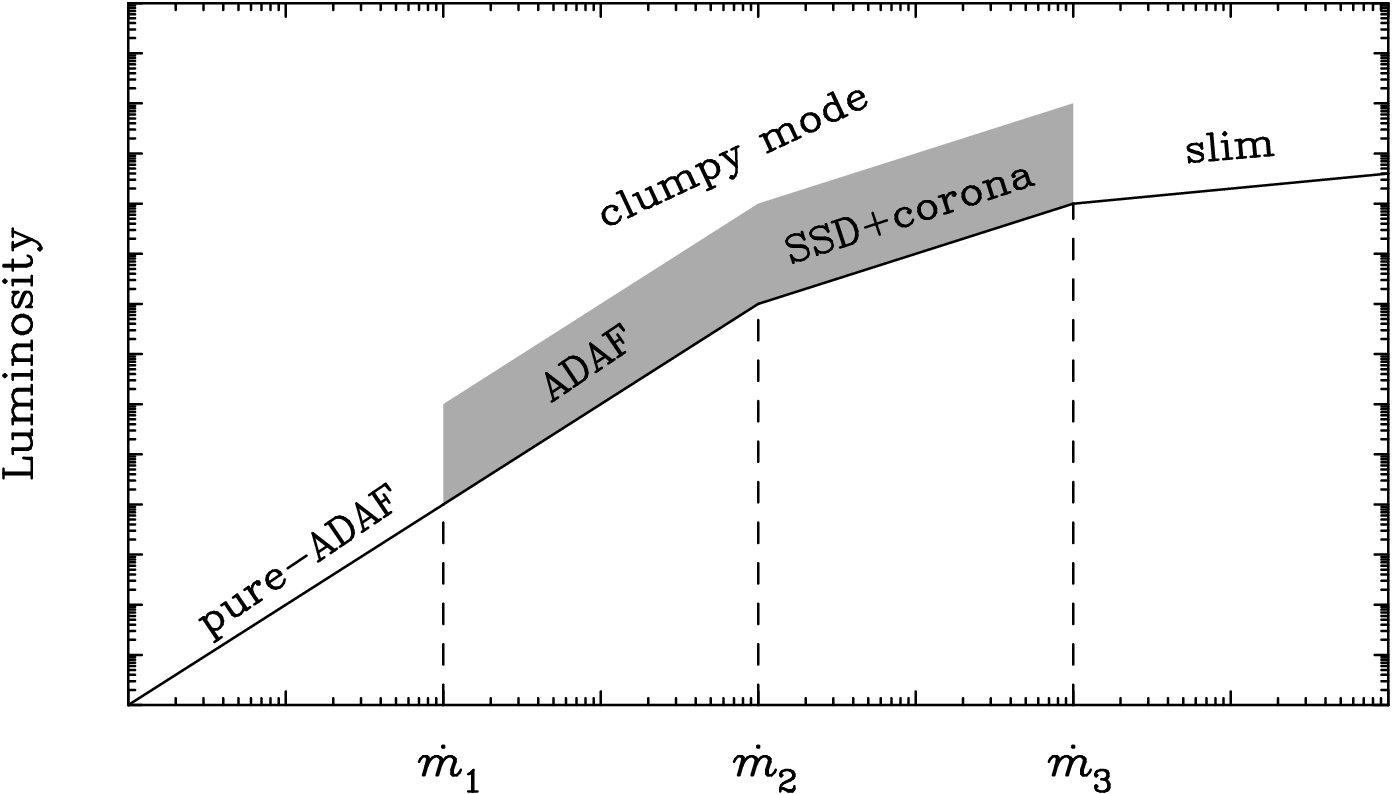}
\figcaption{\footnotesize
Accretion disk models in light of accretion rates. Accretion flows could develop clumpy structure
indicated by the shaded regions. In the ADAF regimes, the radiated luminosity $L\propto \dot{m}^2$ whereas
$L\propto \dot{m}$ in SSD regimes, and $L\propto \ln \dot{m}$ in the slim regime. It is not easy for a
slim disk to develop a clumpy structure since its density is too high to develop thermal instability. On
the other hand, slim disks ($\dot{m}_3\sim 1$) only show weak features of hot corona above the disks,
making the ionization parameter lower than that in SSD. }
\label{disk_regime}
}

The critical accretion rate $\dot{m}_1$ can be roughly estimated from the ionization parameter, which is
defined by $\Xi=L/4\pi R^2c n\kb T$, where $\kb$ is the Boltzmann constant, $L$
is the radiated luminosity, $R$ is the distance to the ionizing source, $n$ and $T$ is the density and
temperature of the cold clumps. The self-similar solution of the simple ADAF model gives the thermal
pressure $P_{\rm A}=1.7\times 10^{16}\alpha^{-1}c_1^{-1}c_3^{1/2}m_{\bullet}^{-1}\dot{m}r^{-5/2}$
${\rm g~cm^{-1}~s^{-2}}$, where $c_1=(5+2\epsilon^\prime)g/3\alpha^2$,
$c_2=\left[2\epsilon^{\prime}(5+2\epsilon^{\prime})g/9\alpha^2\right]^{1/2}$,
$c_3=2(5+2\epsilon^{\prime})g/9\alpha^2$, $g=\left[1+18\alpha^2/(5+2\epsilon^{\prime})^2\right]^{1/2}-1$,
$\epsilon^{\prime}=(5/3-\gamma)/f(\gamma-1)$, $\gamma$ is the adiabatic index and $f$ is the
advection-dominated factor (Narayan \& Yi 1994). In this paper, we use $c_1=0.46$, $c_2=0.48$, $c_3=0.31$
for $\gamma=1.4$, $f=0.9$ unless we point out their specific values\footnote{In the self-similar ADAF model,
$\gamma\in(4/3,5/3)$ depends on the magnetic field density. Actually
the numerical solutions of the ADAF avoid the $\gamma\neq 5/3$ case (Manmoto et al. 1997). For simplicity,
we take an intermediate value of $\gamma=1.4$ in calculations.}. The results
in the present paper are not very sensitive to the values of $\gamma$
and $f$. Using the scaling relation of ADAF, we have its bolometric luminosity
$L_{\rm ADAF}=0.2(\dot{m}/\alpha)^2L_{\rm Edd}$ (Mahadevan 1997). Here we only use the single temperature
of the ADAF model. The ionization parameter defined by $\Xi=L/4\pi R^2 c P_{\rm cl}$ is used to describe
the two-phase medium, where $P_{\rm cl}$ is the internal pressure of the clumps. We assume that the clumps
hold a pressure balance with the ADAF ($P_{\rm cl}=P_{\rm A}$)\footnote{The clumpy-Shakura-Sunyaev
disk shows a different relation of the ionization parameter with the distance to the black hole. Since
radiation pressure dominates in the inner regions, we have $\Xi\propto \dot{m}r^{-1/2}$, where the radiation
pressure $P_{\rm rad}\propto m_{\bullet}^{-1}r^{-3/2}$. Details of clumpy-SSD will be carried out in a
forthcoming paper.}, yielding the critical accretion rate as
\begin{equation}
\dot{m}_1=0.02~\alpha_{0.2}\Xi_{0.1}r_{1000}^{-1/2},
\end{equation}
where $\Xi_{0.1}=\Xi/0.1$, $r_{1000}=R/1000R_{\rm Sch}$, and $R_{\rm Sch}=2.95\times 10^5m_{\bullet}$cm
is the Schwarzschild radius. For cases with $10>\Xi>0.1$ (Krolik 1998), the ionized gas holds a two-phase
state with two different temperatures and a pressure balance between the hot and cold medium. For gas
with $\Xi>10$, only hot phase exists whereas $\Xi<0.1$ only cold phase.
The timescale of the thermal instability is generally given by the line cooling process, which determines
the formation timescale of clumps. With the cooling function (B\"oringher \& Hensler 1989), we have
$\Delta t_{\rm cl}\sim n_e\kb T/n^2\Lambda_{\rm line}=7\times 10^5~n_6^{-1}T_6^{2.3}$s for medium with
one solar abundance. For an ADAF of $1\sunm$ black hole, clumps could be produced at the interacting
regions between the ADAF and the Shakura-Sunyaev disk, namely the evaporation
region, where $\Delta t_{\rm cl}$ will be much shorter than that of the Keplerian rotation. Detailed
analysis is needed to show the production of clumps through thermal instability. Figure 2 shows a
cartoon of the clumpy-ADAF model.

We would like to point out that Kuncic et al. (1996) and Celotti \& Rees (1999) present more arguments
to support the general existence of cold clumps in the accretion disk. This lower limit (equation 1)
for the clumpy-ADAF is only based on the thermal instability and is regarded as a characteristic critical
value. The limit critical accretion rate could get lower if the magneto-rotational instability is included
in the ADAF. Furthermore, we presume that the ADAF part of the global clumpy-ADAF can be described by
the self-similar solution\footnote{The clumps will have dynamical feedback to the ADAF, but also the
reprocessed emission could significantly cool the hot ADAF. We neglect these effects in this paper. Full
treatments of the problem should couple clump equations with the ADAF.}. We do not consider the effects
of ADAF-driven outflow on the clumps as known as advection-dominated inflow and outflows (ADIOs) (Blandford
\& Begelman 1999). However, it would be very interesting to postulate the situation which clumps could
dynamically follow the outflows, forming clumpy outflows. ADIOs with clumps as a potential scenario
will be considered in the future (the referee is acknowledged for this motivated point).

{
\centering
\figurenum{2}
\includegraphics[angle=-90,scale=0.7]{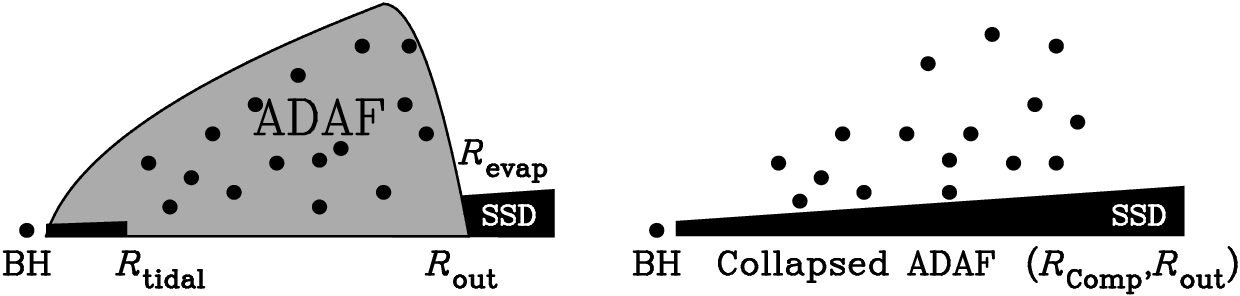}
\figcaption{\footnotesize  Cartoons of clumpy ADAF. {\em Left} panel: A tiny debris disk forms through
accumulation of the tidally disrupted clumps by the black hole within the tidal radius ($R_{\rm tidal}$).
The disk can be accumulated up to the Shakura-Sunyaev regime. We set the tidal radius as the inner edge
of the clumpy-ADAF, which is the outer radius of the debris disk. The radiation from the debris disk has
strong feedback to the hot ADAF through Compton cooling, giving rise to a collapse of the ADAF and quenching
the clumpy accretion. {\em Right} panel: The collapsed ADAF (cADAF). States of clumpy-ADAF transit to a
cADAF through the debris disk, and versus through disk evaporation. This leads to a kind of quasi-periodical
oscillation of the global accretion flows. The fates of clumps remain open, they totally either disappear
or are orbiting around the black hole, or collide with the cADAF. See the text for details. }
\label{disk_model}
}

\subsubsection{Clumps in the clumpy-ADAF}
Existence of cold clumps can be simply justified by the thermal instability. Though detailed analysis
is much beyond the scope of the present paper, we can use the simplified arguments to grasp the
essentials here. The maximum size of clumps is determined by the crossing distance of sound wave
within one Keplerian timescale, otherwise, the clumps are actually like a ring. This yields
$R_{\rm cl}^{\rm max}\sim c_s t_{\rm Kep}\approx 4.4\times 10^{10}~M_8T_4^{1/2}r_{10}^{3/2}$ cm at
radius $R$, where the sound speed
$c_s=10^6~T_4^{1/2}{\rm cm~s^{-1}}$, $r_{10}=R/10R_{\rm Sch}$ and $T_4=T_{\rm cl}/10^4$K. On
the other hand, the minimum size of
clumps is determined by the thermal conduction, below which the clumps will be evaporated. Considering
that line cooling dominates in the cold clumps (with a temperature of $\sim 10^4$K), the cooling rates
$\Lambda_{\rm line}\sim 8.0\times 10^8~n_{14}^2T_4^{-1.3}~{\rm erg~s^{-1}~cm^{-3}}$, where
$n_{14}=n_{\rm cl}/10^{14}~{\rm cm^{-3}}$ (see the Figure 2 in
B\"ohringer \& Hensler 1989). For a Spitzer-like thermal conduction, the conduction rates are of
${\mathscr H}=\nabla\cdot(\kappa_{\rm S}T^{2.5}\nabla T)\sim \kappa_{\rm S}T^{3.5}/R_c^2\approx
5.4\times 10^8~T_{10}^{3.5}/R_{10}^2~{\rm erg~s^{-1}~cm^{-3}}$ (Spitzer 1962). The necessary condition
of ${\mathscr H}\le \Lambda_{\rm line}$ yields
$R_{\rm cl}^{\rm min}\ge 0.82\times 10^{10}~T_{10}^{1.75}n_{14}^{-1}T_4^{0.65}~{\rm cm}$. According
to the pressure balance, the density of clumps is roughly of
$n_{\rm cl}=P_{\rm A}/\kb T_{\rm cl}=2.1\times 10^{16}~M_8^{-1}\dot{m}_{-2}r_1^{-5/2}~{\rm cm^{-3}}$.
Generally, the size and mass of clumps could change with the distance to the black hole. For simplicity,
we use the typical values of cloud mass and radius of clumps:
$m_{\rm cl}=4\pi n_{\rm cl}m_pR_{\rm cl}^3/3\approx  4\times 10^{23}$g and $R_{\rm cl}\sim 10^{11}$cm
for a supermassive black hole with $10^8\sunm$, where $m_p$ is the proton mass.
It should be noted that the Thompson scattering depth of an individual cloud is of
$\tau_{\rm es}=n_{\rm cl}R_{\rm cl}\sigma_{\rm T}=6.65~n_{14}R_{11}$. On the other hand, the temperature
of the clumps could keep a constant about $10^4$K in light of efficient line cooling in the range of
temperature $\sim 10^4$K (Sutherland \& Dopita 1993). Otherwise the clumps will be evaporated by the
surrounding medium. We take an approximately constant temperature of clumps. Table 1 gives the values of
typical clumps for stellar and supermassive black holes.

Turbulence excited by the magneto-rotational instability (MRI) is responsible for transfer of
angular momentum of
the ADAF, and interacts with clumps. However, the MRI-turbulence is not able to destroy the clumps in
light of the energy argument. The energy density of the MRI-turbulence is about $\sim \alpha P_{\rm A}$,
of which $P_{\rm A}$ keeps balance with the thermal energy density of clumps ($P_{\rm cl}$), we have
$\alpha P_{\rm A}<P_{\rm cl}$ since $\alpha<1$. The MRI-turbulence is only a small disturbance to the
clumps. On the other hand, the clump size is much smaller than the typical length of the turbulence
($\alpha H_{\rm A}\sim \alpha R$). Furthermore, the turbulent eddies with a comparable size with
clumps have smaller kinetic energies according to the Kolmogorov's law as $E_{k}\propto k^{-5/3}$ (Landau
\& Lifschitz 1959),
where $E_k$ is the energy per unit wavelength number of turbulence, $k=1/\lambda$ and $\lambda$ is the
length of turbulence. Therefore, the smaller eddies have an energy density $\ll \alpha P_{\rm A}$, which
is only a tiny fraction of clumps. It is thus a good approximation that
clumps are simplified as particles, which can be described by the Boltzmann equation.

Total mass of cold clumps should be self-consistently determined by analysis of global thermal instability,
but, instead, we use the mass ratio defined as $\mathfrak{M}=M_{\rm cl}/M_{\rm ADAF}$ (see Equation 33) as
a free parameter, where $M_{\rm cl}$ is the total mass of clumps and $M_{\rm ADAF}$ is the total mass within
the outer boundary of the ADAF, in the present model. The mass of cold clumps in ADAF is assumed to be
comparable with the ADAF, otherwise, no significant effects can be created (see \S4 for discussions).

When a black hole has relatively low accretion rates, the inner part of the Shakura-Sunyaev disk becomes
optically thin, forming the so-called "hybrid" disk (Shapiro, Lightman \& Eardley 1976; Wandel \& Liang
1991). The ADAF as the inner region of the Shakura-Sunyaev disk with relatively low accretion rates
then develops starting from this radius. Actually, the cold disk will be evaporated by the hot corona,
forming the truncated disk, namely, forming an ADAF starting from the evaporation radius (Meyer \&
Meyer-Hofmeister 1994; Lu et al. 2004), where the evaporation rates are equal to the accretion rates.
In this paper, we take the evaporation radius as the outer boundary radius of the clumpy-ADAF,
$R_{\rm out}=R_{\rm evap}=10^3R_{\rm Sch}$ (Liu \& Taam 2009). The total mass of the ADAF can be
simply estimated by $M_{\rm ADAF}\approx 3.2\times 10^{-2}\sunm~\dot{m}_{-2}M_8r_{1000}^{3/2}$,
where $\dot{m}_{-2}=\dot{m}/10^{-2}$ and $r_{1000}=R/1000R_{\rm Sch}$. Therefore, we have about
${\mathscr{N}}_{\rm tot}={\mathfrak{M}}M_{\rm ADAF}/m_{\rm cl}\sim 6\times 10^9~\mathfrak{M}m_{22}^{-1}$
clumps, indicating that there are a plenty of small dense clumps in the ADAF. The mean distance of clumps
is given by $\langle l\rangle={\mathscr{N}}^{-1/3}\approx 1.0~R_{\rm Sch}$ within $10^3R_{\rm Sch}$,
the crossing timescale $\Delta t_{\rm cross}=\langle l\rangle/\pvvr^{1/2}\approx 10^4M_8r^{-3/2}{\rm s}$
for typical value $\pvvr^{1/2}\approx 0.1c$ (see Figure 4). Comparing with the
Keplerian timescale $t_{\rm K}\approx 10^3M_8r^{-3/2}$s, we find $\Delta t_{\rm cross}>t_{\rm K}$,
indicating that collisions among clumps can be neglected. This guarantees the validity of the colissionless
Boltzmann equation and its moment equations employed in this paper. This could only work for clumpy-ADAF
whereas collisions would be a key mechanism to transport angular momentum outward in a clumpy SSD.

\subsubsection{Interaction between clumps and the ADAF}
Clumps orbiting around the black hole deviate from the dynamics of the ADAF which is a radial flow with
sub-Keplerian rotation. The motion of clumps is controlled by two factors: 1) black hole potential; 2)
drag force arisen by the ADAF. Clumps gain or loss angular momentum by the interaction with the ADAF,
leading to moving
outward or inward, respectively, making the velocity dispersion with ADAF. Angular momentum of clumps is
then carried away by the ADAF, in which $\alpha$-viscosity is responsible for transfer ADAF's
momentum outwards. In this paper, we use the drag force as $F_R=f_R(v_R-V_R)^2$ and
$F_{\phi}=f_{\phi}(v_{\phi}-V_{\phi})^2$, where $V_R$ and $V_{\phi}$ are the radial and $\phi-$velocity
of the ADAF, respectively, $f_R$ and $f_{\phi}$ are two coefficients (Mathews 1990;
Cinzano et al. 1999). In principle, we should use velocities of the MRI-turbulence to estimate
the drag force.

The drag force employed here is an approximation of the drag force in laminar flows. The drag force can
be actually expressed by
$F_{R,\phi}=f_{R,\phi}\left[v_{R,\phi}-(V_{R,\phi}^2+\sigma_{R,\phi}^2)^{1/2}\right]^2
=f_{R,\phi}\left(v_{R,\phi}-\beta_{R,\phi}V_R\right)^{2}$, where
$\beta_{R,\phi}=\left[1+\left(\sigma_{R,\phi}/V_{R,\phi}\right)^2\right]^{1/2}$, $\sigma_{R,\phi}$
are the turbulent
velocities, and the subscripts represent the $R-$ and $\phi-$directions, in turbulent flows. Since
$V_{R,\phi}^2\gtrsim c_s^2=\alpha^{-1}\sigma_{R,\phi}^2$, we have
$\beta_{R,\phi}\approx \left(1+\alpha\right)^{1/2}$.
Considering $\alpha<1$, it would be a good approximation for us to use the laminar drag force
($\beta_{R,\phi}=1$ in this paper).

The two coefficients $f_R$ and $f_{\phi}$ can be approximately taken as constants, which independent
of the ADAF density and the distance to the black hole. Clumps are undergoing contract along with
spiraling-in. For a
simple estimation, we have $n_{\rm cl}R_{\rm cl}^3=n_{\rm cl,0}R_{\rm cl,0}^3$, where the subscript ``0"
indicates the initial value (i.e. at the outer boundary), if individual clumps keep their mass. With the
help of the pressure balance with the ADAF, we have
$R_{\rm cl}=\left(P_{\rm A}/P_{\rm A,0}\right)^{1/3}R_{\rm cl,0}$.
Since the coefficients
$f_{R,\phi}=n_{\rm A}/(n_{\rm cl}R_{\rm cl})=(T_{\rm cl}/T_{\rm A})R_{\rm cl}^{-1}$, we have
$f_{R,\phi}=\left(P_{\rm A}T_{\rm cl,0}/P_{\rm A,0}T_{\rm cl}\right)^{1/3}T_{\rm cl}/T_{\rm A}R_{\rm cl,0}^{-1}$.
Considering the contraction makes clumps a little bit hotter (efficiently being cooled by radiation),
we assume
$T_{\rm cl}\propto R^{-\zeta}$ and $\zeta\ll 1$. We obtain
$f_{R,\phi}\propto R^{(1-4\zeta)/6}R_{\rm cl,0}^{-1}\propto R^{1/30}R_{\rm cl,0}^{-1}$ if $\zeta=0.2$.
Without more details of $\zeta$, we find the coefficients are not sensitive to the distance to the black
holes. We therefore take the coefficients as two constants in this paper.

\subsubsection{Inner edge of the clumpy-ADAF: tidal disruption}
The orbiting clumps are suffering from the tidal disruption governed by the black hole. Self-gravitation
of clumps is negligible, however, clumps still survive through keeping a pressure balance with the
surroundings if the tidal distortion can be overcome by the thermal pressure of the ADAF. The tidal force
reads $F_{\rm tidal}\approx G\mbh m_{\rm cl} R_{\rm cl}/R^3$ for a cloud, where $G$ is the gravitational
constant and $R$ is the distance to the hole. This tidal force is balanced by the thermal pressure of
the ADAF, namely, $F_{\rm tidal}=4\pi P_{\rm A}R_{\rm cl}^2$.
Since the clumps keep a pressure balance with the ADAF, we have $P_{\rm A}=n_{\rm cl}\kb T_{\rm cl}$, where
$n_{\rm cl}=m_{\rm cl}/\frac{4\pi}{3} R_{\rm cl}^3m_p$ is the mass density of the clumps.
Tidal disruption happens when $F_{\rm tidal}\ge 4\pi n_{\rm cl}\kb T_{\rm cl} R_{\rm cl}^2$,
yielding a natural limit of the inner boundary radius
\begin{equation}
{\cal R}_{\rm in}=\left(\frac{G\mbh m_pR_{\rm cl}^2}{4\pi \kb T_{\rm cl}}\right)^{1/3}
                 \approx \left\{\begin{array}{l}
                 8.0 ~M_8^{-2/3}R_{11}^{2/3}T_4^{-1/3}R_{\rm Sch},\\
                                             \\
                 8.0 ~M_1^{-2/3}R_{3}^{2/3}T_4^{-1/3}R_{\rm Sch}\end{array}\right.
\end{equation}
where $M_8=\mbh/10^8\sunm$, $M_1=\mbh/1\sunm$, $R_{11}=R_{\rm cl}/10^{11}$cm, $R_{3}=R_{\rm cl}/10^3$cm
and $T_4=T_{\rm cl}/10^4$K. Here we set the temperature of clumps $T_{\rm cl}=10^4$K. Within the tidal
radius, clumps are destroyed to form a disk of debris or mixed with the ADAF. Some papers have studied
the formation of accretion
disk after tidal disruption of stars by supermassive black holes (e.g. Cannizzo et al. 1990; Strubbe \&
Quataert 2009). It is not the main goal of investigating the detailed processes of destroyed clumps to
form an accretion disk, however, we focus on its influence on the ADAF in this paper. Following the
popular treatment, we assume that the tidal radius is the location of disk after the
orbit of captured debris has been circularized.

We would stress here that we use the parameters of clumps at the tidal radius throughout the whole
ADAF though clumps are undergoing contraction. Collisions are neglected for the clumpy-ADAF, thus the
contraction is not important in the present situation. We use the values of the parameters listed in Table
1 throughout the ADAF. Clumps could be different in size and density in outer parts of the ADAF.
However, it will be very important for the standard disk with clumps. In such a case, clumps merger and
fragment through collisions, depending on the size of the clumps. This is much beyond the scope of the present
paper.

\subsection{Collisionless Boltzmann equation}
Defining the distribution function as
${\mathscr F}=\Delta {\cal N}/R\Delta R\Delta z\Delta \phi\Delta\textbf{v}$,
dynamics of clumps can be generally described by the Boltzmann equation.
Unlike the normal stellar system, the clumps are moving in the SMBH potential, but also dragged by the
ADAF. We start from the origin of collisionless Boltzmann equation (4-11) in Binney \& Tremaine (1987)
\begin{equation}
\frac{\pf}{\pp t}+\sum_{\alpha=1}^6\frac{\pp \left(\mathf\dot{w}_{\alpha}\right)}{\pp w_{\alpha}}=0,
\end{equation}
where \textbf{(x,v)}$\equiv$ \textbf{w} and ${\bf \dot{w}\equiv (\dot{x},\dot{v}}$) are the coordinates
in the
phase space. Generally, for a stellar system, $\sum_{\alpha=1}^6\pp\dot{w}_{\alpha}/\pp w_{\alpha}=0$ holds.
The Boltzmann equation reduces to $\pf/\pp t+\sum_{\alpha=1}^6\dot{w}_{\alpha}\pf/\pp w_{\alpha}=0$.
The drag force on the clumps depends on the velocity, making the Boltzmann equation complicated. Considering
the dependence of acceleration of clumps on their velocity, we re-cast the Boltzmann equation
\begin{equation}
\frac{\pf}{\pp t}+\sum_{i=1}^3\left(v_i\frac{\pf}{\pp x_i}+\dot{v}_i\frac{\pf}{\pp v_i}+
\mathf\frac{\pp \dot{v}_i}{\pp v_i}\right)=0.
\end{equation}
The third term in the bracket arises from the drag force, which disappears in a conservative system as
described by Equation (4-13a) in Binney \& Tremaine (1987). In a cylindric coordinate, we have
\begin{equation}
\frac{\pf}{\pp t}+\dot{R}\frac{\pf}{\pp R}+\dot{\phi}\frac{\pf}{\pp \phi}+\dot{z}\frac{\pf}{\pp z}+
\dot{v}_R\frac{\pf}{\pp v_R}+\dot{v}_{\phi}\frac{\pf}{\pp v_{\phi}}+\dot{v}_z\frac{\pf}{\pp v_z}+
\mathf\left(\frac{\pp \dot{v}_R}{\pp v_R}+\frac{\pp \dot{v}_{\phi}}{\pp v_{\phi}}+
\frac{\pp \dot{v}_z}{\pp v_z}\right)=0,
\end{equation}
where $\dot{R}=v_R$, $\dot{\phi}=v_{\phi}/R$ and $\dot{z}=v_z$. Motion equations of an individual cloud
read
\begin{equation}
\dot{v}_R=-\frac{\pp \Phi}{\pp R}+\frac{v_{\phi}^2}{R}+\FR;~~~~
\dot{v}_{\phi}=-\frac{1}{R}\frac{\pp \Phi}{\pp \phi}-\frac{v_Rv_{\phi}}{R}+\Fphi;~~~~
\dot{v}_z=-\frac{\pp \Phi}{\pp z},
\end{equation}
where $\Phi=G\mbh/(R^2+z^2)^{1/2}$,
$\FR=f_R(\vr-\Vr)^2$ and $\Fphi=f_{\phi}(\vphi-\Vphi)^2$ are the drag forces per unit mass in
the $R-$ and $\phi-$direction, respectively, $\Vr$ and $\Vphi$ are the radial and
the rotational velocities of the ADAF. It should be noted
that $f_R$ and $f_{\phi}$ might be functions of cloud's parameter, such as, radius and density.
Here we neglect the drag forces in $z-$direction. We have
$\sum_{\alpha=1}^6\pp\dot{w}_{\alpha}/\pp w_{\alpha}=2f_{\phi}(\vphi-\Vphi)+2f_R(\vr-\Vr)$ in
the cylindric coordinate frame. Considering the $\phi-$symmetry of the clumpy disk, we have
\begin{equation}
\begin{array}{rr}
{\displaystyle
\frac{\pf}{\pp t}+v_R\frac{\pf}{\pp R}+{v_z}\frac{\pf}{\pp z}+
\left(\frac{v_\phi^2}{R}-\frac{\pp \Phi}{\pp R}+\FR\right)\frac{\pf}{\pp v_R}+
\left(\Fphi-\frac{v_Rv_{\phi}}{R}\right)\frac{\pf}{\pp v_{\phi}}-
\frac{\pp \Phi}{\pp z}\frac{\pf}{\pp v_z}}+ & \\
                                            & \\
{\displaystyle 2\mathf\left[f_\phi\left(\vphi-\Vphi\right)+f_R\left(\vr-\Vr\right)\right]} &= 0.
\end{array}
\end{equation}
This is the final version of the Boltzmann equation used for dynamics of clumps in the following
sections.

\subsection{Moment equations}
Following the popular treatment, we solve the moment equations of the Boltzmann equation. We define
the averaged parameter in velocity-space as $\langle X\rangle=\mathn^{-1}\int X\mathf d\vec{v}$, where
$X$ is a parameter. Here we use $\mathn=\int {\cal N}d\vec{v}$.
The zeroth-order moment equation can be obtained by integrating Equation (7)
\begin{equation}
\frac{\pp \mathn}{\pp t}+\frac{1}{R}\frac{\pp}{\pp R} \left(R\mathn \pvr\right)
+\frac{\pp}{\pp z}\left(\mathn\pvz\right)=0.
\end{equation}
The first-order moment equations can be gained through multiplying Equation (7) by
$v_R$, $v_{\phi}$ and $v_z$, respectively, and integrating the equations in velocity-space.
The first moment equation is given by
\begin{equation}
\begin{array}{rr}
{\displaystyle \frac{\pp}{\pp t}\left(\mathn\pvr\right)+\frac{\pp}{\pp R}\left(\mathn\pvvr\right)+
\frac{\pp}{\pp z}\left(\mathn\pvrvz\right)+\mathn\left(\frac{\pvvr-\pvvphi}{R}+
\frac{\pp \Phi}{\pp R}\right)- }& \\
                                &  \\
{\displaystyle \mathn f_R\left(\pvvr-2\pvr\Vr+\Vr^2\right)}&=0,
\end{array}
\end{equation}
the second
\begin{equation}
\frac{\pp}{\pp t}\left(\mathn\pvphi\right)+\frac{1}{R^2}\frac{\pp}{\pp R}\left(R^2\mathn\pvrphi\right)+
\frac{\pp}{\pp z}\left(\mathn\pvphiz\right)-\mathn f_{\phi}\left(\pvvphi-2\pvphi\Vphi+\Vphi^2\right)=0,
\end{equation}
and the third
\begin{equation}
\frac{\pp}{\pp t}\left(\mathn\pvz\right)+\frac{\pp}{\pp R}\left(\mathn\pvrvz\right)+
\frac{\pp}{\pp z}\left(\mathn\pvvz\right)+\frac{1}{R}\mathn\pvrvz+\frac{\pp\Phi}{\pp z}\mathn=0.
\end{equation}
We will use these moment equations to discuss the dynamics of the clumpy disk.

For the $\phi-$ and $z-$direction symmetric clumpy ADAF, we have $\pvz=0$, $\pvrvz=0$, $\pvphiz=0$ and
re-cast the continuity equation as
\begin{equation}
\frac{1}{R}\frac{\pp}{\pp R} \left(R\mathn \pvr\right)=0,
\end{equation}
yielding the accretion rates of clumps as $\dot{M}_{\rm cl}=-2\pi R H_{\rm A}\mathn m_{\rm cl}\pvr$.
The first moment equation is reduced to
\begin{equation}
\frac{\pp}{\pp R}\left(\mathn\pvvr\right)+
\mathn\left(\frac{\pvvr-\pvvphi}{R}+
\frac{\pp \Phi}{\pp R}\right)-\mathn f_R\left(\pvvr-2\pvr\Vr+\Vr^2\right)=0,
\end{equation}
the second to
\begin{equation}
\frac{1}{R^2}\frac{\pp}{\pp R}\left(R^2\mathn\pvrphi\right)-
\mathn f_{\phi}\left(\pvvphi-2\pvphi\Vphi+\Vphi^2\right)=0,
\end{equation}
and the third to
\begin{equation}
\frac{\pp}{\pp z}\left(\mathn\pvvz\right)+
\frac{\pp\Phi}{\pp z}\mathn=0.
\end{equation}

Table 2 gives a summary of input and output parameters used in the present model.
The above moment equations describe the dynamics of clumps, however, these are not close.
We have to supplement additional physical considerations to proceed. We distinguish
two classes of the dynamics in light of the strength of coupling between clumps and ADAF.
When $f_R$ and $f_\phi$ are large enough, the clumps are strongly coupled with ADAF so that
the averaged dynamics of clumps follows the ADAF. For small $f_R$ and $f_\phi$, the clumps
are weekly coupled with the ADAF, showing weak dependence on the ADAF.

\section{Structure of clumpy disk}
\subsection{Strong-coupling case}
In the strong-coupling case, the drag force is so strong that the averaged dynamics of the clumps
follow the ADAF, namely, $\pvr=\Vr$ and $\pvphi=R\Omega_{\rm A}$, where $\Omega_{\rm A}$ is rotational
velocity of the ADAF. So the factor $f_\phi$ should be larger than a critical one. This can be understood
by the fact that both the specific angular momentum and the kinetic energy of clumps are about
same with the gas of ADAF since they are born in the ADAF. 
Therefore, we have
\begin{equation}
\pvr=-2.12\times 10^{10}\alpha c_1r^{-1/2}~{\rm cm~s^{-1}},
\end{equation}
\begin{equation}
\pvphi=2.12\times 10^{10}c_2r^{-1/2}~{\rm cm~s^{-1}},
\end{equation}
$\pvr/\pvphi=-\alpha c_1/c_2$, and the height of the clumpy disk
\begin{equation}
H_{\rm A}=2.95\times 10^5c_3^{1/2}c_2^{-1}mr~{\rm cm}.
\end{equation}
From the continuity equation, we have
\begin{equation}
\mathn=-\frac{\dot{M}_{\rm cl}}{4\pi RH_{\rm A}\pvr m_{\rm cl}},
\end{equation}
where $\dot{M}_{\rm cl}$ is the averaged accretion rates of clumps. From $\phi-$motion equation,
we have
\begin{equation}
\pvvphi=\frac{1}{\mathn f_{\phi}}\frac{1}{R^2}\frac{d}{dR}\left(R^2\mathn\pvrphi\right)+2\pvphi\Vphi-\Vphi^2
       =-\frac{1}{2}\frac{\pvrphi}{f_{\phi}R}+2\pvphi\Vphi-\Vphi^2,
\end{equation}
and $\pvvphi/\pvphi^2=1+\alpha c_1/2c_2f_{\phi}R$. The radial motion is rewritten by
\begin{equation}
\frac{d\pvvr}{dR}+\frac{1}{R}\left(1+\frac{d\ln\mathn}{d\ln R}-f_RR\right)\pvvr+
\left(\left\langle\frac{\pp\Phi}{\pp R}\right\rangle-\frac{1}{R}\pvvphi+f_R\Vr^2\right)=0,
\end{equation}
where
$$
\left\langle\frac{\pp\Phi}{\pp R}\right\rangle=\frac{1}{H_{\rm A}}\int_0^{H_{\rm A}}\frac{\pp\Phi}{\pp R}dz
\approx \frac{G\mbh}{R^2}.
$$
Considering $d\ln \mathn/d\ln R=-3/2$ from Equation (19), we have
\begin{equation}
\frac{d\pvvr}{dR}-\frac{1}{R}\left(f_RR+\frac{1}{2}\right)\pvvr+R\left(\Omega_{\rm K}^2-
\frac{\pvvphi}{\pvphi^2}\Omega_{\rm A}^2\right)+f_R\Vr^2=0.
\end{equation}
Clearly, the $\phi-$motion has strong influence on the radial motion. Inserting $\pvvphi$ and $\pvphi$, we
have
\begin{equation}
\frac{d\pvvr}{dR}-\frac{1}{R}\left(f_RR+\frac{1}{2}\right)\pvvr+\left[(1-c_2^2)R-
\frac{1}{2}\alpha c_1c_2f_{\phi}^{-1}\right]\Omega_{\rm K}^2+f_R\Vr^2=0.
\end{equation}
With the outer boundary condition of $\pvvr=V_{\rm out}^2$ at $R=R_{\rm out}$, we have the solution as
\begin{equation}
\pvvr=c^2\left\{\frac{1}{2}\left[\alpha^2c_1^2\Gamma_R\Lambda_{\frac{3}{2}}(\Gamma_R,r)
       +(1-c_2^2)\Lambda_{\frac{5}{2}}(\Gamma_R,r)
       -\frac{\alpha c_1c_2}{2\Gamma_{\phi}}\Lambda_{\frac{7}{2}}(\Gamma_R,r)\right]
       +\frac{V_{\rm out}^2}{c^2r_{\rm out}^{1/2}}e^{-\Gamma_Rr_{\rm out}}\right\}r^{1/2}e^{\Gamma_R r},
\end{equation}
where $\Gamma_R=f_RR_{\rm Sch}$, $\Gamma_{\phi}=f_{\phi}R_{\rm Sch}$ and the function
$$\Lambda_{q}(\Gamma_R,r)=\int_r^{r_{\rm out}}x^{-q}e^{-\Gamma_Rx}dx,$$
and $q=3/2,5/2,7/2$. Figure 3 shows the properties of the function $\Lambda_q$.
Equation (24) gives the solution of the clumps, which deviates from the ADAF.

{
\centering
\figurenum{3}
\includegraphics[angle=-90,scale=0.45]{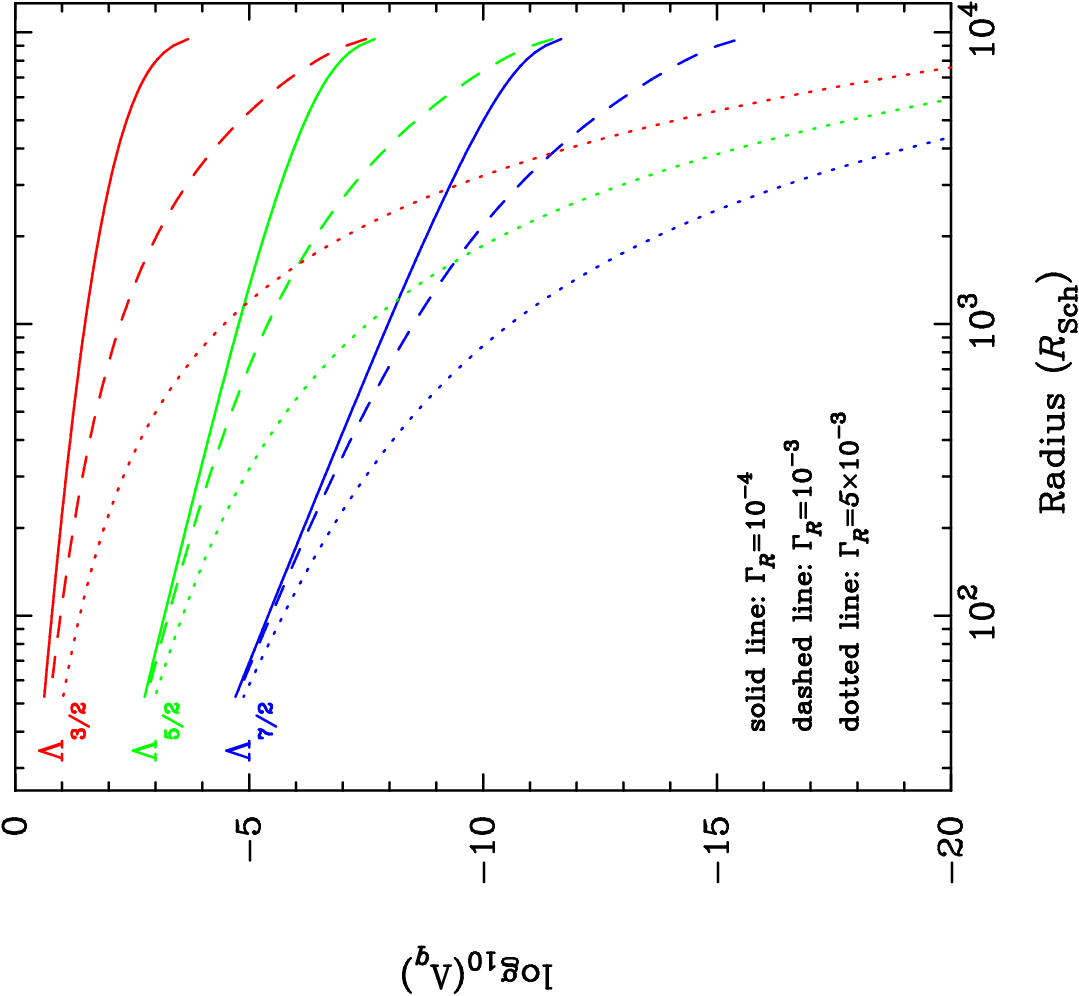}
\figcaption{\footnotesize Properties of the function $\Lambda_q$. It is found that $\Lambda_q$ is
sensitive to the parameter $\Gamma_R$, but only at large radius. This property has strong influence
on the radial velocity dispersion at large radii rather than that at small ones.}
\label{sdss_spectra}
}

Physical meanings of each terms in Equation (24) can be examined under some extreme cases.
Generally, clumps are controlled by two factors: 1) black hole potential; 2) $\phi-$ and
$R-$direction drag forces. For a drag-free cloud, its orbit is determined purely by
the black hole. When $f_R$ tends to zero, namely $\Gamma_R=0$, clumps are orbiting around black
hole with $\phi-$drag. Angular momentum of clumps is transferred by the ADAF, in which the
popular $\alpha-$prescription works for outward transportation of the angular momentum of the
ADAF, giving rise to fast spiral down to the black hole. In such a case, we have
\begin{equation}
\pvvr= \frac{1}{3}(1-c_2^2)\frac{c^2}{r}\left[1-\left(\frac{r}{r_{\rm out}}\right)^{3/2}\right]-
      \frac{\alpha c_1c_2}{20\Gamma_{\phi}}\frac{c^2}{r^2}
      \left[1-\left(\frac{r}{r_{\rm out}}\right)^{5/2}\right]+
      \left(\frac{r}{r_{\rm out}}\right)^{1/2}V_{\rm out}^2.
\end{equation}
The first term is the orbital motion around the black hole. The second term results in the accordance
of cloud motion with the ADAF. The $\phi-$drag decreases the velocity dispersion between the clumps
and the ADAF. When the $\phi-$drag is strong enough, we have the critical value
\begin{equation}
\Gamma_{\phi}^c\approx \frac{3\alpha c_1c_2}{10\left(2-2c_2^2-3\alpha^2 c_1^2\right)r_{\rm in}}
               \approx 8.7\times 10^{-4}~\alpha_{0.2}r_{10}^{-1}.
\end{equation}
Here we neglect
the terms of $(r_{\rm in}/r_{\rm out})$. When $\Gamma_{\phi}=\Gamma_{\phi}^c$, coupling with the ADAF
is so strong that the velocity dispersion of clumps with the ADAF is zero at the tidal capture radius
($R_{\rm in}$), namely, $\pvvr^{1/2}=V_{\rm in}$ at $R=R_{\rm in}$, where $V_{\rm in}$ is the radial
velocity of the ADAF. So the strong coupling is referred to the case with $\Gamma_{\phi}>\Gamma_{\phi}^c$.
Models with $\Gamma_{\phi}<\Gamma_{\phi}^c$ are the weak coupling and the strong coupling approximation
does not work. We will discuss the case below.

{
\centering
\figurenum{4}
\includegraphics[angle=-90,scale=0.8]{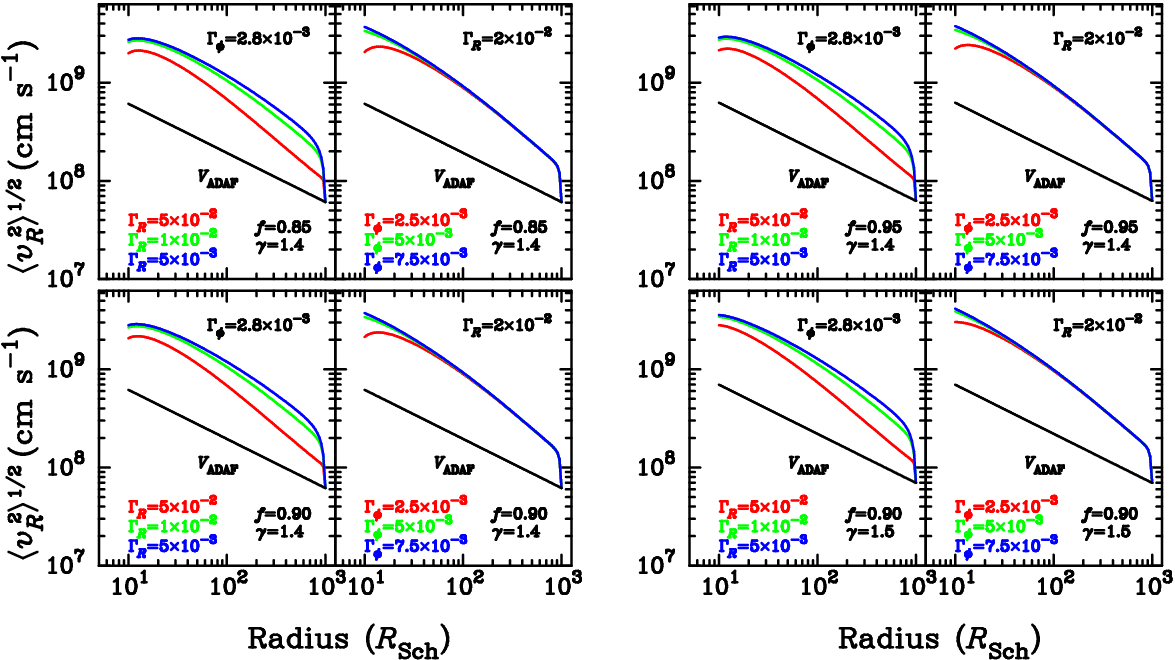}
\figcaption{\footnotesize Solution of the clumpy-ADAF disk. The value of $\pvvr^{1/2}$
could be 10 times the radial velocity of ADAF. We note that the outer boundary does not
significantly affect the properties of the inner part of clumpy-ADAF. The four panels show the
dependence of the solution on the two index $\gamma$ and factor $f$. It is found that the results
are not sensitive to the two constants $\gamma$ and $f$.}
\label{sdss_spectra}
}

For an extremely strong-coupling, namely $\Gamma_{\phi}\rightarrow \infty$, clumps tend to have
\begin{equation}
\begin{array}{rl}
\pvvr=&\displaystyle{c^2\left\{\frac{1}{2}\left[\alpha^2c_1^2\Gamma_R\Lambda_{\frac{3}{2}}(\Gamma_R,r)
       +(1-c_2^2)\Lambda_{\frac{5}{2}}(\Gamma_R,r)\right]
       +\frac{V_{\rm out}^2}{c^2r_{\rm out}^{1/2}}e^{-\Gamma_Rr_{\rm out}}\right\}r^{1/2}e^{\Gamma_R r}},\\
       &         \\
\approx &\displaystyle{\frac{1}{3}(1-c_2^2)\frac{c^2}{r}\left[1-
\left(\frac{r}{r_{\rm out}}\right)^{3/2}\right]+\left(\frac{r}{r_{\rm out}}\right)^{1/2}V_{\rm out}^2,}
\end{array}
\end{equation}
namely, reaches its maximum. Here the first term with $\Lambda_{\frac{3}{2}}$ is
always smaller than the others.

Figure 4 shows the solutions of the clumpy-disk for different parameters of the drag forces. For
fixed $\Gamma_{\phi}$ case, it can be found that the radial drag strongly influences the velocity
dispersion of clumps at large radii. On the other hand, for fixed $\Gamma_R$ cases, $\pvvr$ is
mainly determined by the $\Gamma_{\phi}$ and reaches its maximum as shown by Figure 4. We
find the term involving $\Lambda_{3/2}$ is always smaller than the other two in Equation (24).
This is due to radial velocity of the ADAF is smaller than the rotational, being represented by
multiplying the factor $\alpha$. Though the influence of the radial drag can be neglected, the
present treatments are complete. The most important is that $\pvvr^{1/2}\sim 10 \pvr$ for the
extremely strong coupling case from Figure 4. This means that the accretion rates of clumps are
actually enhanced.

\subsection{Weak-coupling case}
When the birth of clumps are not very tightly linked with the ADAF in dynamics, the strong-coupling
between clumps and the ADAF is relaxed. In such a case, of weak coupling with the ADAF, the assumptions
of $\pvr=\Vr$ and $\pvphi=\Vphi$ do not work, and cloud dynamics resembles the stellar system.
We introduce the velocity dispersions: $\sigmar^2=\pvvr-\pvr^2$, $\sigmaphi^2=\pvvphi-\pvphi^2$ and
$\sigma_z^2=\pvvz-\pvz^2$ to solve the moment equations. The moment equations should be closed
up by physical considerations. Similar to stellar dynamics, we assume $\sigma_R^2=k_R\sigma_z^2$
and $\sigma_{\phi}^2=k_{\phi}\sigma_z^2$, where $k_R$ and $k_{\phi}$ are two constants.
Since clumps are produced by the ADAF in light of thermal instability
and the velocity dispersion of clumps should be less than the sound speed of the ADAF,
the vertical height of the clumpy disk does not exceed the
ADAF height, we assume $H=H_{\rm A}$. Employing $\pvz=0$ and $\sigma_z^2=\pvvz$, we have
\begin{equation}
\sigma_z=H_{\rm A}\Omega_{\rm K},
\end{equation}
After some algebraic manipulations, we re-cast Equation (17) and (18)
\begin{equation}
\left(\frac{\pvr^2-\sigma_R^2}{\pvr}\right)\frac{d\pvr}{dR}-
     \frac{\pvphi^2+\sigma_{\phi}^2}{R}+\frac{d\sigma_R^2}{dR}+
     \left\langle\frac{\pp \Phi}{\pp R}\right\rangle=0,
\end{equation}
and
\begin{equation}
\pvr\frac{d\pvphi}{dR}+\frac{\pvr}{R}\pvphi-f_{\phi}\left[\sigma_{\phi}^2+
               \left(\pvphi-V_{\phi}\right)^2\right]=0,
\end{equation}
where Equation (19) is used.

For a weak-coupling case, clumps are mainly orbiting around the black hole. The
first term with $d\pvr/dR$, $\sigma_{\phi}^2/R$ and $d\sigma_R^2/dR$ are of the same order,
but much smaller than the $\pvphi^2/R$ and $\pp \Phi/\pp R$. So we have
\begin{equation}
\pvphi\approx -\left\langle\frac{\pp \Phi}{\pp R}\right\rangle=V_{\rm K},
\end{equation}
where $V_{\rm K}$ is the Keplerian velocity. Since $\pvphi\approx V_{\rm K}$, we have
$\sigma_{\phi}\approx 0$ and $dv_{\phi}/dR\approx -v_{\phi}/2R$, Equations (30) yields
\begin{equation}
\pvr=\frac{2f_{\phi}R}{V_{\rm K}}\left(V_{\rm K}-V_{\phi}\right)^2
    =7.0\times 10^4~\Gamma_{-5}r^{1/2}~{\rm cm~s^{-1}},
\end{equation}
where $\Gamma_{-5}=\Gamma_{\phi}/10^{-5}$. The mean velocity of clumps linearly proportional to the
factor $f_{\phi}$. The velocity dispersion $\pvvr=\sigma_R^2+\pvr^2$ and
$\pvvphi=\sigma_{\phi}^2+\pvphi^2$ are known if $k_R$ and $k_{\phi}$ are known, however, the present
model is not able to determine $k_R$ and $k_{\phi}$ in a self-consistent way. The two parameters should
be constrained by observations. The weak coupling case does not have significant observable effects,
we therefore remain it here.

As a brief summary, dynamics of clumps is very different from the ADAF whatever for strong and weak coupling
cases between the clumps and the ADAF, but the dynamical properties of clumps depends on the ADAF. The
prominent properties are: the velocity dispersion of clumps is one order higher or much smaller than the
radial velocity of the ADAF in strong and weak-coupling cases, respectively. For the weak coupling,
solutions of the clumps weakly coupled with the ADAF show that their properties are similar to stars
in the black hole potential field. These properties of strong-coupling clumpy-ADAF determine variabilities
of accretion disk of black holes.

Finally, we would like to point out the roles of magnetic fields discussed by Kuncic et al. (1996). For magnetized
clumps, magnetic fields might be is in equipartition with the thermal pressure. The magnetic field plays
a role in resisting the MRI-turbulence and makes the existence of clumps more persistent. Furthermore,
magnetic fields lower the thermal conductivity (Spitzer 1962), and thus make clumps survive more robust.
The real situations could be much complicated. Future numerical simulations would uncover more details
of roles of magnetic fields in the clumpy-ADAF.

\section{Radiation: observational appearances}
Radiative properties of clumps in AGNs and X-ray binaries have been extensively studied by several authors
(Kuncic et al. 1997; Celotti \& Rees 1999; Malzac \& Celotti 2002; Merloni et al. 2006). Much attention is
given to the reprocessed emission from the clumps, emission lines are the main features. In such a case,
profiles of emission lines could be broadened by the motions of clumps, and can be calculated
through the method of Whittle \& Saslaw (1986). The present paper investigates the case in which the
debris of the tidally disrupted clumps is accumulating with time until a transient disk of the debris around
the black hole forms. We show that the debris disk plays a key role in the radiation of the global accretion
flows, in particular, feedback to the ADAF driven by the debris disk leads to a kind of quasi-periodical
oscillation of the flows.

\subsection{Capture rates of clumps}
Considering the mass of clumps $\Delta M_{\rm cl}=4\pi RH_{\rm A}\mathn m_c \Delta R$ within $R$ to
$R+\Delta R$ whereas the gas mass of the ADAF is $\Delta M_{\rm A}=4\pi RH_{\rm A}\rho_{\rm A}\Delta R$,
we find the mass fraction of the clumps to the ADAF
\begin{equation}
\mathfrakm=\frac{\Delta M_{\rm cl}}{\Delta M_{\rm A}}=\frac{\dot{M}_{\rm cl}}{\dot{M}_{\rm A}},
\end{equation}
where $\dot{M}_{\rm A}$ is the accretion rates of the ADAF. In principle, the parameter $\mathfrakm$
should be determined self-consistently by the model of cloud formation, but this is beyond the scope
of the present paper. We treat it as a free parameter in the model.

The capture rates are given by the numbers of clumps which entre the tidal radius per unity time.
Considering an interval time of $\Delta t$, we have the number of the captured clumps
through the surface of the tidal radius
$\Delta \mathn=2\pi R_{\rm in}H_{\rm in}\mathn_{\rm in}\pvvr^{1/2}\Delta t$, $H_{\rm in}$ and
$\mathn_{\rm in}$ are the height of clumpy disk and clump number density at the tidal radius, respectively.
Therefore we have the capture rates as
$\dot{\mathscr R}=\lim_{\Delta t\rightarrow 0}\left(\Delta \mathn/\Delta t\right)$
\begin{equation}
\dot{\mathscr R}=\delta \mathfrakm\dot{m}_{\rm A}\frac{\dot{M}_{\rm Edd}}{m_{\rm cl}}
                \approx\left\{\begin{array}{l}
                1.4\times 10^7~\eta_{0.1}^{-1}\delta_1\mathfrakm\dot{m}_{-2}m_{10}^{-1}M_1~{\rm s^{-1}},\\
                                            \\
                3.6\times 10^9~\eta_{0.1}^{-1}\delta_1\mathfrakm\dot{m}_{-2}m_{22}^{-1}M_8~{\rm month^{-1}},
                \end{array}\right.
\end{equation}
where $\delta=\pvvr^{1/2}/\pvr$, $\delta_1=\delta/10$, $\dot{m}_{-2}=\dot{M}_{\rm A}/10^{-2}\dot{M}_{\rm Edd}$.
We note that the capture rates are very high since $\pvvr^{1/2}\sim 10\pvr$.

Since the capture timescale $\dot{\mathscr{R}}^{-1}$ is much smaller than the accretion of the debris onto black
holes, the captured clumps are monotonically accumulating with time. For a single capture event, the tidally
disrupted debris has complicated fates, which could resemble the capture stars (e.g. Rees 1988). However,
it should be noted the difference of the present case from the captured star: the debris of disrupted
clumps will interact with the ADAF. The circularization timescale could be of multiple Keplerian timescale of
orbiting the black hole, leading to mix with local ADAF. On the other hand, after a series of captured clumps,
a tiny disk of debris will be formed within the timescale of $\Delta t_{\rm cir}$ if the clumps take kinetic
energy enough. The interaction among the debris of the captured clumps is very complicated, however, it is
reasonable to assume that the timescale of forming the debris disk is of the order of $\Delta t_{\rm cir}$.

\subsection{Fates of captured clumps}
Fates of disrupted clumps depend on both cloud's properties and the ADAF density. Similar to the case
of the captured star, the debris of a disrupted cloud has a specific kinetic energy
$\varepsilon \sim 3(G\mbh/R_P)(R_{\rm cl}/R_P)$ for clumps approaching the black hole with a parabolic
orbit, where $R_P$ is the pericenter distance of the parabolic orbit (Lacy 1982; Evans \& Kochanek 1989).
After multiple cycles of the parabolic orbits, the gas flow will be circularized, forming a disk. The
fallback timescale is given by
\begin{equation}
\Delta t_{\rm cir}^{\rm min}\approx\left\{ \begin{array}{l}
                  0.7~M_1^{5/2}R_{_{P,3}}^3R_{\rm cl, 3}^{-3/2}~{\rm s},\\
                                     \\
                  2.2~M_8^{5/2}R_{_{P,3}}^3R_{\rm cl, 11}^{-3/2}~{\rm yr},
                  \end{array}\right.
\end{equation}
where $R_{_{P,3}}=R_{_P}/3R_{\rm Sch}$ (e.g. Strubbe \& Quataert 2009). Since the debris of clumps
is embedded in the ADAF, the interaction between them is unavoidable. This makes it possible for the
debris to mix with the ADAF when the interaction timescale is shorter than that of the circularization,
otherwise the debris will form a disk of its own, called as a debris disk. This timescale is actually
for the accumulation of debris disk ($\Delta t_{\rm acc}\sim \Delta t_{\rm cir}^{\rm min}$).

\subsubsection{Mixed with the ADAF}
The swept mass rates of the debris is of $\dot{\mathscr{M}}\sim \pi R_{\rm cl}^2 n_{\rm A}\pvvr^{1/2}m_p$
and the lost energy rates are of $\dot{E}_{\rm K}\sim \dot{\mathscr{M}}c_s^2$, where $c_s$ is the sound speed
of the ADAF. The timescale of dissipating the kinetic energy of the debris is given by
\begin{equation}
\Delta t_{\rm diss}=\frac{E_{\rm K}}{\dot{E}_{\rm K}}\sim \frac{m_p\pvvr R_{\rm cl}}{\kb T_{\rm cl}\pvr}
                   \approx\left\{\begin{array}{l}
                   1.2~\delta_1\pvvr_{8}^{1/2}R_3T_4^{-1} ~{\rm s},\\
                                      \\
                   3.8~\delta_1\pvvr_{8}^{1/2}R_{11}T_4^{-1}~{\rm yr},
                   \end{array}\right.
\end{equation}
where $\pvvr_8^{1/2}=\pvvr^{1/2}/10^8{\rm cm~s^{-1}}$, and the approximation of the pressure balance
between clumps and the ADAF is used. We find
that $\Delta t_{\rm diss}\sim \Delta t_{\rm cir}^{\min}$ generally holds for small $\pvvr$ cases, but
$\Delta t_{\rm diss}$ is significantly longer than $\Delta t_{\rm cir}^{\rm min}$ for large $\pvvr$
cases. This indicates formation of a debris disk, especially, for the very strong-coupling case.

When the debris is mixed with the ADAF, the density of the ADAF within the tidal radius will be
enhanced generally, resulting in that cooling of the ADAF increases. Cooling enhancement by the
mixture with the captured clumps could be moderately in this case. We will not pay much attention
on this case, instead, on the case of forming a debris disk.

\subsubsection{Formation of debris disk}
As we shown below, capture
of clumps is much faster than accretion onto black holes, yielding accumulation of debris around the hole.
Since the clumps orbits are in the ADAF, the new disk of the accumulated debris will be inside the ADAF.
Detailed formation of the debris disk could be very complicated (more than the case of the tidal disrupted
stars, see Rees 1988; Strubbe et al. 2009). Here we assume that the formed disk holds the approximation of
radiation-pressure dominated regions of the Shakura-Sunyaev disk since most of the gravitational energy
will be released in this region. The accretion timescale driven by viscosity is given by
$t_{\rm debris}\approx R_{\rm debris}^2/\nu=\alpha^{-1}\left(H_{\rm debris}/R\right)^{-2}t_{\rm K}$, where
$t_{\rm K}=1/\Omega_{\rm K}$ is the Keplerian rotation timescale, $R_{\rm debris}$ and $H_{\rm debris}$ are
the typical radius and scale height of the debris disk, respectively, and $\nu$ is the kinetic viscosity.
Using the SSD solution, we have the accretion timescale
\begin{equation}
\Delta t_{\rm debris}=\left\{\begin{array}{l}
                   1.56~\alpha_{0.2}^{-1}M_1\dot{m}_{0.1}^{-2}r_{10}^{7/2}~{\rm s},\\
                                                                \\
                   4.96~\alpha_{0.2}^{-1}M_8\dot{m}_{0.1}^{-2}r_{10}^{7/2}~{\rm yr},
                   \end{array}\right.
\end{equation}
where $r_{10}=R_{\rm in}/10R_{\rm Sch}$ and $\alpha_{0.2}=\alpha_{\rm SSD}/0.2$. We find that
$\dot{\mathscr{R}}^{-1}\ll \Delta t_{\rm debris}$. Accumulation of the debris
around the black hole follows the capture of clumps.

For an interval $\Delta t$, the accumulated mass of the captured clumps is
$\Delta M_{\rm cl}=m_{\rm cl}\dot{\mathscr {R}}\Delta t$, where we neglect the swallowed clumps by
the black hole during the accumulation. We have the accretion rates of the debris disk
as $\dot{M}_{\rm debris}=\Delta M_{\rm cl}/\Delta t_{\rm debris}$, and the dimensionless rate is
\begin{equation}
\dot{m}_{\rm debris}
                    =\delta\mathfrak{M}\dot{m}_{\rm A}\left(\frac{\Delta t}{\Delta t_{\rm debris}}\right)
                    =0.1~\delta_1{\mathfrak{M}}\dot{m}_{-2}\Delta t_1,
\end{equation}
where $\Delta t_1=\Delta t/1.0\Delta t_{\rm debris}$. We stress that the accretion rates of the debris
disk is higher by one order than the undergoing ADAF arise from the fact of the radial velocity of clumps
($\delta\sim 10$) for the strong coupling case. We would like to point out that the storage of the debris
is undergoing through capturing the clumps since it carries too much kinetic energy to be directly accreted
or mixed with the local ADAF.

It should be pointed out that the above estimation is based on the radiation pressure-dominated
solution of SS73 model. The transition radius from gas to radiation pressure-dominated regions is
$R_{\rm tr}/R_{\rm Sch}\approx 104.1\left(\alpha m_{\bullet}\right)^{2/21}\left(\dot{m}/\eta_{0.1}\right)^{16/21}$.
For $\dot{m}=0.1$, we find that $R_{\rm tr}\approx 18.0\left(\alpha m_{\bullet}\right)^{2/21}R_{\rm Sch}>{\cal R}_{\rm in}$
holds generally. This means that the debris disk should be generally radiation pressure-dominated. For
those debris disks dominated by gas pressure could still be in ADAF-regime until they reach in the SSD regimes.
The present estimations are valid.

\subsection{Feedback: collapse of ADAF?}
\subsubsection{Compton cooling as feedback to the ADAF}
We show that a debris disk forms within $\Delta t\sim \Delta t_{\rm debris}$ in the regime
of the Shakura-Sunyaev model. The disk is radiating at a quite large luminosity,
$L_{\rm debris}=\eta \dot{M}_{\rm debris}c^2\sim 10^{45}\dot{m}_{0.1}M_8$\ergs, where
$\dot{m}_{0.1}=\dot{m}_{\rm debris}/0.1$. Photons from the debris disk
spanning from optics to UV for supermassive black holes and from UV to $\lesssim 1.0$keV
for a few solar mass black holes provide extra sources to cool the hot electrons in
the ADAF through Compton cooling with a timescale
\begin{equation}
\Delta t_{\rm Comp}=\frac{n_e\kb T}{\Lambda_{\rm Comp}}=\left\{\begin{array}{l}
                    54.1~M_1\dot{m}_{0.1}^{-1}r_{1000}^2~{\rm ms}\\
                                            \\
                    2.1~M_8\dot{m}_{0.1}^{-1}r_{1000}^2~{\rm month},
                    \end{array}\right.
\end{equation}
where $\Lambda_{\rm Comp}=2n_e\kb T\sigma_{\rm T}F_{\rm debris}/m_ec^2$ is the Compton cooling
rates, the energy flux from the debris disk is $F_{\rm debris}=L_{\rm debris}/4\pi R^2$,
$L_{\rm debris}=\dot{m}_{\rm debris}L_{\rm Edd}$. Setting $t_{\rm Comp}=\Delta t_{\rm debris}$,
we have the Compton radius, within which the ADAF is driven by the emergent photons from the
debris disk to collapse through Compton cooling
\begin{equation}
R_{\rm Comp}=5447.0~\alpha_{0.2}^{-1/2}r_{10}^{7/4}\dot{m}_{0.1}^{-1/2}~R_{\rm Sch}.
\end{equation}
We find that $R_{\rm Comp}>(R_{\rm evap},R_{\rm out})$, suggesting that the global ADAF will be
cooled through Compton cooling.
With such a strong feedback\footnote{The steady ADAF is formed by the balance between gravity heating and
cooling (free-free, synchrotron and inverse Compton scattering). Since the photon fluxes from
the debris disk are much larger than the ADAF-generated energy flux
($L_{\rm debris}\sim 0.1L_{\rm Edd}$ whereas gravity heating $L_{\rm G}\lesssim 10^{-2}L_{\rm Edd}$
in the ADAF itself), the ADAF is rapidly cooled through the Compton cooling without sufficient
heating of released gravitational energy. Esin (1997) discussed the influence of non-local radiation
on the ADAF, but it is different from the present case.}, the ADAF collapses into
geometrically thin disk since it has angular momentum. The collapse timescale is mainly controlled by
the vertical gravity of the black hole. The collapsing velocity is given by
$v_{\rm ff}=H_{\rm Comp}\Omega_{\rm K}$, and the time scale is
$\Delta t_{\rm collapse}=\Omega_{\rm K}^{-1}=0.44~r_{1000}^{3/2}M_1~{\rm s}=1.7~r_{1000}^{3/2}M_8~{\rm yr}$.
This timescale is much shorter than the dynamical, shocks could be thus formed during the
collapse and could heat the collapsed ADAF (hereafter cADAF). In this paper, we neglect this heating,
which could be balanced by cooling of being condensed gas. The collapse stops until the cADAF reaches
a new dynamical equilibrium of the Shakura-Sunyaev disk with a scale height of
$H_{\rm cADAF}/R=4\times 10^{-3}\dot{m}_{-2}r^{-1}$. In such a case clumps are then orbiting around
the black holes without the ADAF-driven drag if they are bounded by magnetic field. It is also plausible
for clumps collide with the cold disk and then are captured by the disk. They could undergo fast expansion
and totally disappear since the pressure balance is broken. This feedback gives rise to quenching the
clumpy accretions. The debris disk is playing as a switch in these processes.

\subsubsection{Collapsed ADAF and revived clumpy-ADAF}
The cADAF is undergoing two processes: 1) itself proceeds accretion onto black
holes at a viscosity timescale; 2) it may be evaporated by hot corona from outer to inner regions. Though
the formation of hot corona on the Shakura-Sunyaev disk remains open, we presume here that the two competing
processes determine the post appearance of the cADAF. For the cADAF as an
Shakura-Sunyaev disk, it is still radiation-pressure dominated and has a timescale of
\begin{equation}
\Delta t_{\rm cADAF}=\left\{\begin{array}{l}
                     0.5~\alpha_{0.2}^{-1}M_1\dot{m}_{0.1}^{-2}r_{1000}^{7/2}~{\rm yr},\\
                                           \\
                     5.0\times 10^7~\alpha_{0.2}^{-1}M_8\dot{m}_{0.1}^{-2}r_{1000}^{7/2}~{\rm yr},
                     \end{array}\right.
\end{equation}
which is much longer than $\Delta t_{\rm collapse}$. This makes the cADAF has the accretion
rates of the previous ADAF, and radiate at
$L_{\rm cADAF}=\dot{m}_{\rm A}L_{\rm Edd}\approx 1.25\times 10^{44}\eta_{0.1}\dot{m}_{-2}M_8$
\ergs. The disk enters a relatively brighter state than the ADAF. The rising time scale is about
$\Delta t_{\rm debris}+\Delta t_{\rm Comp}$ from the ADAF state. However, the fate of the cADAF
is determined by the competition between the fueling black hole and evaporating the cADAF.

According to numerical calculations (Liu \& Taam 2009), the evaporation timescale is given by
\begin{equation}
\Delta t_{\rm evap}=\frac{M_{\rm cADAF}}{\dot{M}_{\rm evap}}
                 \approx 1.45~r_{1000}^{3/2}\dot{m}_{-2}\dot{m}_{\rm evap,-2}^{-1}~{\rm yr},
\end{equation}
where $M_{\rm cADAF}$ is the mass of the collapsed ADAF, which is roughly equal to total mass of
the ADAF within  $R_{\rm out}$, and $\dot{M}_{\rm evap}$ is evaporation rates. The estimation simply
follows from $\dot{M}_{\rm evap}\sim 10^{-2}\dot{M}_{\rm Edd}$, which depends on viscosity somehow.
We would stress here that the evaporation timescale does not depend on black hole mass. Comparing
$\Delta t_{\rm cADAF}$ with $\Delta t_{\rm evap}$, we have
$\Delta t_{\rm tran}=\min(\Delta t_{\rm cADAF},\Delta t_{\rm evap})$, indicating that evaporation
could govern the post appearance of the cADAF in AGNs whereas viscosity of accretion
does in X-ray binaries. We point out that accretion onto black holes in AGNs and X-ray binaries
are different in this way. After the interval of $\Delta t_{\rm tran}$, the clumpy-ADAF revives.
A new cycle starts.

Figure 5 shows a cartoon of a cycle of the state transition in a black hole clumpy-ADAF.
After the time $\Delta t_{\rm tran}$, a new ADAF develops, and the object enters a low/hard state,
switching on the clumpy accretion. Clumpy-ADAF is at low/hard state, but the cADAF corresponds
to the high/soft state since it is Shakura-Sunyaev disk. State transition happens through the feedback
of transient disk of the debris. The collapsed ADAF powering the high state will be brighter than the
ADAF at the low/hard state. These processes correspond to transition of states in black hole X-ray
binaries. For low luminosity AGNs, there could be a component as a big blue bump in some Low Ionization
Nuclear Emission Regions (LINERs), especially in some LINERs with broad components of emission lines
(Ho 2008). Furthermore, BL Lac objects have ADAF and some of them show light curves with
quasi-periodical modulations, which could be explained by the present model. We stress that the
transient disk plays a key role in feedback to the ADAF. The
present model predicts a transition of accretion flows. There are two processes
of shorter bursts: 1) emission from the debris disk; 2) cooling of the ADAF
through the inverse Compton scattering. The first outburst proceeds to the second, especially,
it is in soft band, and the second is a burst in hard X-ray band. Detailed comparison
with observations of X-ray binaries and AGNs would determine the size of the cADAF and the disk
of the cloud's debris.

{
\centering
\figurenum{5}
\includegraphics[angle=-90,scale=0.5]{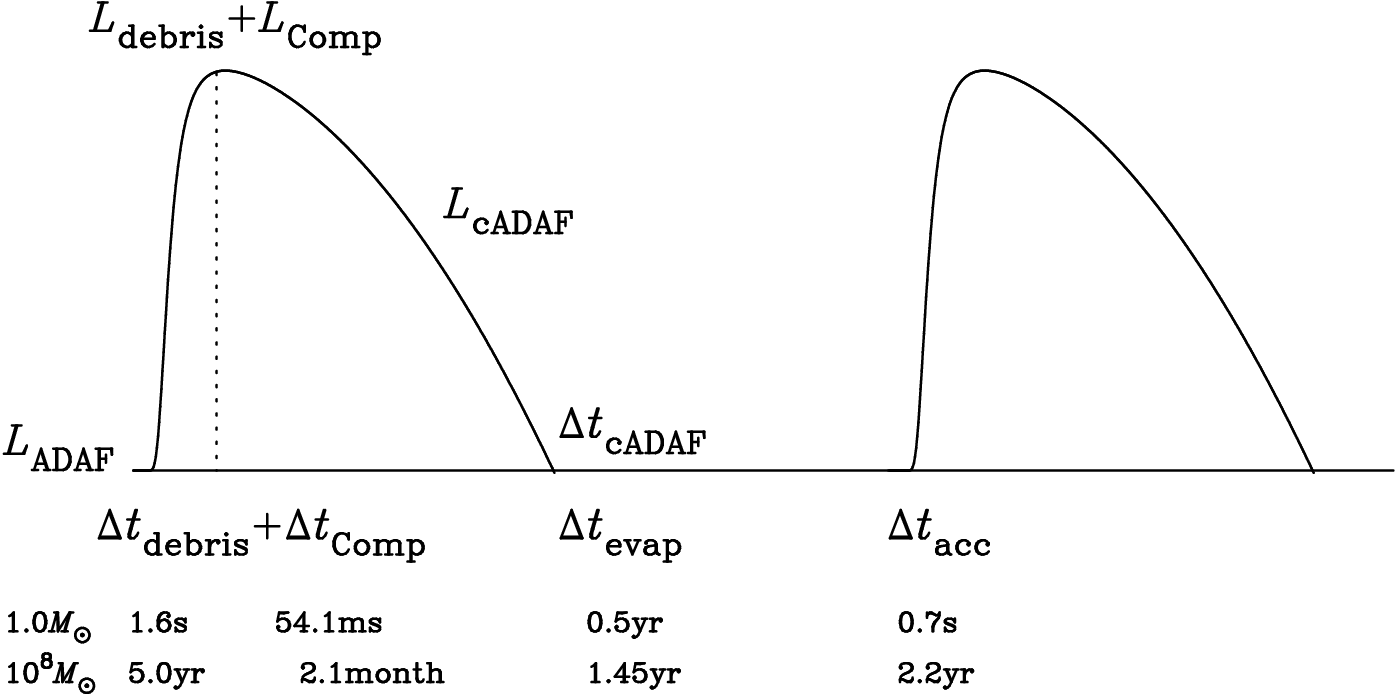}
\figcaption{\footnotesize Characterized light curves of clumpy-ADAF. The long term averaged luminosity
radiates at $L_{\rm ADAF}$. The debris disk of tidally disrupted clumps is in Shakura-Sunyaev regime
and efficiently radiates at a higher luminosity, and drives the hot ADAF to collapse into geometrically
thin. The cADAF brings the accreting black hole to high/soft state with luminosity $L_{\rm cADAF}$.
The accretion timescale of the debirs disk ($\Delta t_{\rm debris}$) determines the size of the cADAF.
Gas of the cADAF proceeds to being accreted onto black hole, in the meanwhile, evaporation develops
a new ADAF with a timescale of $\Delta t_{\rm evap}$ governing the recurrence of the clumpy-ADAF in AGNs.
After $\min(\Delta t_{\rm cADAF},\Delta t_{\rm evap})$, clumpy accretion revives. This leads to a state
transition in X-ray binaries, and sets up a quasi-periodical light curve in low luminosity AGNs. The
quasi-period is give by $\Delta t_{\rm acc}$ and a burst duration by
$\Delta t_{\rm debris}+\Delta t_{\rm Comp}+\Delta t_{\rm cADAF}$. It should be noted that the timescales
labeled are characterized ones and depends on some parameters. The timescales in the cartoon are not
scaled. }
\label{sdss_spectra}
}

\subsection{Discussions}
When an accretion flow has low enough rates, it turns to an ADAF by evaporation (Meyer \& Meyer-Hofmeister
1994) and would become clumpy-ADAF with a critical accretion rate of $\dot{m}\gtrsim 0.02\alpha_{0.1}$
as shown in Figure 1. Clumps are finally disrupted by the tidal force of the black hole,
forming a transient tiny disk. Liu et al. (2007) and Meyer-Hofmeister \& Meyer (2011) suggest that an
inner disk may be formed through condensations of ADAF if the accretion rates are in a reasonable range.
In principle, this condensation model results from thermal instability, and is physically equivalent
to the clumpy-ADAF model suggested in this paper. However, their model is different from the
present in three aspects: 1) the tiny disk is persistent in Liu et al. (2007), but it is transient in
the present model; 2) only two components (cold disk and ADAF) in Liu et al. (2007) and many clumps
in the present model; 3) dynamics is different. As we shown below, the present model can conveniently
explain more observations, besides the observed iron K$\alpha$ lines in AGNs (Meyer-Hofmeister \&
Meyer 2011), such as variabilities of X-ray binaries and radio-loud AGNs.

In the present model, we assume that clumps in the ADAF keep constant mass and radius before being tidally
disrupted. Despite of these assumptions, it holds the main features of the clumpy accretion. We may relax
some assumption in future studies. Here, we omit to discuss the weak coupling case of the clumpy-ADAF
because it may
give results without significant difference from the pure ADAF. On the other hand, if the accretion rates are
in Shakura-Sunyaev disk regime, the dynamics of clumps could be changed into a phase driven by collisions
among the clumps as well as the drag force in $R-$ and $\phi-$direction. Observations show that X-ray regions
are partially covered by clumps (Gallo et al. 2004; Ballantyne et al. 2004; Ricci et al. 2010; see a review
of Turner \& Miller 2009), implying clumps in the clumpy-SS disk will be much larger than the present. These
contents will be discussed as the main goals of the second papers. We treat the outer boundary as
the evaporation radius, where production of clumps is likely happening due to the thermal instability here.
This should be issued in more detailed in a future paper.

Finally, numerical simulations of the dynamics and radiation of the clumpy-ADAF are worth doing to give
more details like in dusty torus (Stalevski et al. 2011). We briefly discuss the radiation from the
clumpy-ADAF in light of time scales of the undergoing processes. Collapse of the ADAF deals with energy
release of gravitational energy in the vertical direction though this energy is smaller than the debris
disk. The cADAF radiates is simplified as a Shakura-Sunyaev disk. This is valid when the collapse
time scale is much shorter than the accretion time scale. Future numerical simulations will uncover the
full processes of the collapse of the ADAF and radiation from the collapsed ADAF.

\section{Observational tests and applications}
The present clumpy-ADAF model only applies to those black holes which have a steady accretion rates
in the ADAF regime with $\dot{m}\gtrsim 10^{-2}$ as shown by Equation (1). These accreting black holes
could be used to test the predictions of the present model. We briefly illustrate to apply the present
model to X-ray binaries and low luminosity AGNs to potentially explain the related phenomena, but
detailed applications will be given in a separated paper.

\subsection{Low luminosity AGNs: LINERs and BL Lacs}
The active galactic nuclei well-known in the ADAF regime are LINERs (e.g. Ho 2008) and some BL Lac
objects (e.g. Wang et al. 2002; 2003). Since the timescale of the transient disk is much longer than
that in X-ray binaries, the component of the transient disk will be observed easily, but
it varies at a time scale of years.

{\em LINERs}: evidence has been found for presence of significant component of the big blue bump
observed in normal AGNs and quasars (Maoz et al. 2007). Pure reprocessing emission from clumps
(their Figure 1 in Celotti \& Rees 1999) is not enough to explain the component since the reprocessing
emission gets a peak at $10^{14.5}$Hz. The transient disk originated from the captured clumps definitely
contributes to its emission to $10^{14}\sim 10^{16}$Hz in light of the Shakura-Sunyaev disk model. It
is trivial to test this model in LINERs since the component of the debris disk has a variability
with a time scale of years depending on the black hole masses and accretion rates of the ADAF. It
should be noted that the selected LINERs to test the model should have relative higher accretion
rates in the clumpy-ADAF regime rather than the pure ADAF mode. These LINERs have broad H$\alpha$
components, which are $25\%$ LINERs (Ho 2008). They could have relatively higher accretion rates
(Elitzur \& Ho 2009) and thus contain clumps in the ADAF. These LINERs are expected to be monitored
for variabilities to test the accretion processes.

{\em BL Lac} objects: they have lower accretion rates, likely in ADAF mode (Wang, Ho \& Staubert 2002;
2003; Barth
et al. 2003). It is expected that some of them could have clumpy-ADAF. The observed emission is overwhelmed
by the boosted emission of relativistic jet, and the emission from the debris disk or cADAF is not directly
visible. It is generally postulated that the ADAF will produce a relativistic jet somehow (e.g. Meier 2001),
evidenced by X-ray binaries (e.g. Fender et al. 2010), for example, the radio galaxy 3C 120 (Marscher et al.
2002) and 3C 111 (Tombesi et al. 2011). As we have shown, the presence of the debris disk drives
disappearance of the ADAF through efficient Compton cooling, and results in quenching jet formation. In such
a case, thermal components ($L_{\rm debris}$, $L_{\rm Comp}$ and $L_{\rm cADAF}$) could be observed if the
relativistic boosting jet emission less overwhelms this component. More interestingly, the model predicts
a periodical presence of the clumpy-ADAF which could lead to a light curves
with quasi-periodical modulation in light of intermittent production of jet in the clumpy-ADAF mode.
Jet production quenches in the period of $\Delta t_{\rm debris}+\Delta t_{\rm Comp}+\Delta t_{\rm cADAF}$.
Radio light curves from the website\footnote{http://www.astro.lsa.umich.edu/obs/radiotel/umrao.php}
with quasi-periodical modulations of a few years observed in some BL Lac objects and radio galaxies could be
explained by the present model, for example, 0235+164, 3C 120,
0607-157, 0727-115, 1127-145, 1156+295, 1308+326, 1335-127, OT 129, BL Lac, 3C 446, 3C 345. These objects
have supermassive black holes with mass of $10^8\sim 10^9\sunm$ (Ghisellini \& Celotti 2001;
Barth et al. 2003; Wang et al. 2002, 2003)
and Eddington ratios of $10^{-3}\sim 10^{-2}$ (Wang et al. 2003).

We note that some sources from this website, such as, 3C 273 are different from the sources above
mentioned. These sources may have higher accretion rates than the ADAF in BL Lac types sources. This
implies the necessity of clumpy standard accretion disk discussed in previous sections.

\subsection{Black hole X-ray binaries: state transition}
Extensive reviews of observed properties of black hole X-ray binaries and related theoretical explanations
have been given by Remillard \& McClintock (2006), McClinktock \& Remillard (2006), Done et al. (2007)
and Belloni et al. (2011). Some black hole X-ray binaries show repeat transition from low to high states
and versus. The present model predicts repeat transitions. From the long term light curves
of 4U 1630-47, XTE J1650-500, XTE J1720-318, H 1743-322, SLX 1746-331, XTE J1859 and Cyg X-1,
they usually have Eddington ratio of $L/L_{\rm Edd}\sim 10^{-2}$, but show variabilities with
two orders (Done et al. 2007). They show bursts with very steep
rising and slow declining. For a simple estimation at
hard state, the hard X-ray luminosity (at $\sim 100$keV) from the ADAF
$L_{\rm HX}\sim 2.5\times 10^{35}\left(\mbh/\sunm\right)\dot{m}_{-2}$\ergs\ (Mahadevan 1997). Once
formation of the debris disk, it is radiating at a luminosity of
$L_{\rm SX}\sim 1.3\times 10^{37}\left(\mbh/\sunm\right)\left(\dot{m}_{\rm debris}/0.1\right)$ and
the accreting black hole transit to at a high state with a around two order change (
$L_{\rm HX}/L_{\rm SX}\sim 10^{-2}$). Detailed calculations of state transition and variability
will be carried out in a forthcoming paper.

On the other hand, the presence of clumps in ADAF will increase the radiative efficiency
of the ADAF, appearing the hard states with high luminosity (e.g. Figure 5 in the most recent review of
Belloni et al. 2011). After the ADAF collapses, jet production quenches and the accretion flows then transit
to high state. We note that the collapse of the ADAF will squeeze itself. The squeezed gas could be blown
away by the radiation, yielding outflows. This happens during the transition of states, especially from
low to high states. Detailed comparisons with observations will be given in a future paper.

\section{Conclusions}
We show that an ADAF becomes a clumpy ADAF composed of cold clumps arising from thermal instability
when $\dot{m}\gtrsim 0.02\alpha_{0.2}r_{1000}^{-1/2}$.
We set up the dynamics of clumpy accretion onto black holes, and focus on the clumpy ADAF in this paper.
This model fills the regimes of the accretion mode from standard disks to pure ADAF.
Angular momentum of clumps is transported by the ADAF. We discuss the strong and weak coupling cases
separately. The inner edge of the clumpy disk is set at the radius of tidal disruption. Analytical
solutions of the clumpy ADAF are obtained for the two cases. For the strong coupling case, the
root of the averaged radial velocity square can be one order higher than the ADAF, resulting in a fast
capture of clumps through tidal force. For weak coupling case, clumps are mainly orbiting around the
black hole. The tidally disrupted clumps are accumulating with time until an efficiently radiating disk
forms. The ADAF is driven to collapse through the Compton cooling of photons from the disk. As a consequence,
clumpy accretion stops then. This constitutes a quasi-periodical modulation. The model could explain the
state transition in X-ray binaries as well as the broad K$\alpha$ lines in their low states. Thermal
components observed in LINERs can be in principle explained by the model. Moreover, the observed light
curves with quasi-periodical modulations in some BL Lac objects can be explained by the theoretical model
as a result of quenching the ADAF through Compton cooling driven by the transient disk formed by the
tidally captured clumps.

We study a simplified model that could capture much interesting $-$ though not all of the detailed physics
of the clumpy-ADAF. We would stress that the present model neglects collisions among clumps, but they
are important for the clumpy Shakura-Sunyaev disk. Moreover, productions of cold clumps could happen in
the entire regions where radiation pressure dominates and source function of clumps will appear in the
Boltzmann equation. The complicated properties of variabilities are arisen by the collisions of clumps
and will be studied in a forthcoming paper.

\acknowledgements{The authors are very grateful to the anonymous referee for useful reports, which clarify
some points in the early version of the paper. We appreciate the stimulating discussions among the members
of IHEP AGN group. S. Mineshige is thanked for motivated discussions as to the clumpy-ADAF during his stay
of visiting IHEP. L. C. Ho and H.-Y. Zhou are acknowledged for useful comments and suggestions of LINER
variabilities. The research is supported by NSFC-10733010, -10821061, and -11173023 and 973 project
(2009CB824800). }

\clearpage

\begin{deluxetable}{ccccccc}
\tabletypesize{\small}
\tablewidth{0pt}
\tablecaption{\small {\sc Values of Cold Cloud Parameters in the Clumpy-disk}}
\tablehead{
\multicolumn{3}{c}{stellar mass BH}& &\multicolumn{3}{c}{supermassive BH}\\ \cline{1-3}\cline{5-7}
$R_{\rm cl}~({\rm cm})$ & $m_{\rm cl}~({\rm g})$ & $n_{\rm cl}~({\rm cm^{-3}})$ &
& $R_{\rm cl}~({\rm cm})$ & $m_{\rm cl}~({\rm g})$ & $n_{\rm cl}~({\rm cm^{-3}})$}
\startdata
$10^3$  & $4\times 10^{10}$ & $10^{23}$& & $10^{11}$ & $4\times 10^{23}$ & $10^{14}$
\enddata
\end{deluxetable}

\begin{deluxetable}{ll}
\tabletypesize{\small}
\tablewidth{0pt}
\tablecaption{\small {\sc Parameters of the Present Model}}
\tablehead{Parameter & Physical meanings}
\startdata
input parameters    &                    \\ \hline
$\mbh$              & black hole mass\\
$\dot{M}_{\rm A}$   & accretion rates of the continuous flow\\
$\dot{M}_{\rm cl}$  & averaged accretion rates of the clumps\\
$m_{\rm cl}$        & averaged mass of individual clumps \\
$R_{\rm in}$        & the tidal radius as the inner radius of the clumpy-ADAF\\
$R_{\rm out}$       & outer radius of the clumpy-ADAF\\
$R_{\rm cl}$        & averaged radius of individual clumps\\
$n_{\rm cl}$        & hydrogen number density of clumps\\
$\mathfrakm$        & ratio of the ADAF and clumps mass ($=\dot{M}_{\rm cl}/\dot{M}_{\rm A}$) \\
$\alpha$            & viscosity parameter of the ADAF\\
$\Gamma_R$          & coefficient of $R-$direction drag force\\
$\Gamma_{\phi}$     & coefficient of $\phi-$direction drag force\\ \hline
output parameters   &                    \\ \hline
$\mathn$            & number density of the clumps\\
$\pvr$              & averaged $R-$direction velocity\\
$\pvvr$             & averaged of $v_R^2$\\
$\pvphi$            & averaged $\phi-$direction velocity\\
$\pvvphi$           & averaged values of $v_{\phi}^2$\\
$\dot{\mathscr{R}}$ & capture rates of clumps
\enddata
\end{deluxetable}

\clearpage

\appendix
\section{Energy equation of clumps: the second moment equation}
Section 2.2 gives the zeroth- and the first-order moment equations of the clumps. We use the
strong-coupling condition to close up the moment equations. Energy equation of clumps can
be obtained from the second-order moment equation, which can be obtained by multiplying and integrating
$v_{_R}v_{\phi} d\vec{v}$:
\def\la{\langle}
\def\ra{\rangle}
\def\ds{\displaystyle}
\begin{equation}
\begin{array}{l}
	\ds \frac{\pp}{\pp R} \left( \mathn \la v_{_R}^{2}v_{\phi} \ra\right)
	+\frac{\pp}{\pp z} \left( \mathn \la v_{_R}v_{\phi}v_{z} \ra\right) 	
	-\frac{\mathn \la v_{\phi}^{3} \ra}{R}+\mathn\pvphi \frac{\pp \Phi  }{\pp R}
	+\frac{2 \mathn \la v_{_R}^{2}v_{\phi} \ra}{R}-f_{_R} \mathn \la v_{\phi}v_{_R}^{2} \ra\\
                        \\
	\ds +2f_{_R}\mathn V_{_R}\la v_{_R}v_{\phi} \ra	-f_{_R}\mathn V_{_R}^{2}  \la v_{\phi} \ra
	-f_{\phi}  \mathn \la v_{_R}v_{\phi}^{2} \ra +2f_{\phi}\mathn V_{\phi} \la v_{_R}v_{\phi} \ra
	-f_{\phi}\mathn V_{\phi}^{2} \la v_{_R} \ra=0.
\end{array}
\end{equation}
For the $\phi-$ and $z-$direction symmetric clumpy ADAF, we have the following relations
\begin{equation}	
\la v_{_R}^{2}\left(v_{\phi}-\pvphi\right)\ra=\la v_{_R}^{2}v_{\phi}\ra-\pvvr\pvphi=0,
\end{equation}
\begin{equation}	
\la v_{_R}v_{z}\left(v_{\phi}-\pvphi\right)\ra=\la v_{_R}v_{z}v_{\phi}\ra-\la v_{_R}v_{z}\ra\pvphi=0,
\end{equation}
\begin{equation}
\la \left(\vphi-\pvphi\right)^{3}\ra=\la v_{\phi}^{3}\ra-3\pvvphi\pvphi+2\pvphi^{3}=0,
\end{equation}
\begin{equation}	
\la v_{_R}v_{z} \ra = 0,
\end{equation}
and subtract $\vphi$ times the moment equation (9), then we obtain the energy equation
\begin{equation}	
\frac{\pp \ln \left( \pvphi R \right) }{\pp \ln R}\frac{\mathn \pvvr\pvphi}{R} 	
-\frac{2\mathn  \pvphi \sigma_{\phi}^{2}}{R}-f_{\phi}  \mathn \la v_{_R}v_{\phi}^{2} \ra
+2f_{\phi}\mathn V_{\phi} \la v_{_R}v_{\phi} \ra-f_{\phi}\mathn V_{\phi}^{2}
\la v_{_R} \ra =0,
\end{equation}
where
\begin{equation}
	\sigma_{\phi}^{2} = \pvvphi - \pvphi^{2}.
\end{equation}
When $f_{\phi}=0$, equation (A6) reduces to equation (4-51) in Binney \& Tremaine (1987).
The energy equation introduces a new unknown parameter $\la \vr\vphi^2\ra$. In order to
close up the moment equations, observational relations should be employed, such as,
ratios of velocity dispersion. This can be done as in galactic dynamics, but it is very
difficult for the present case. We thus take the strong coupling approximations.

\end{document}